\documentclass[showpacs,preprintnumbers,amsmath,amssymb,twocolumn,superscriptaddress,prb]{revtex4}
\usepackage{graphicx}
\usepackage{amsfonts}
\usepackage{amsmath, amsthm, amssymb}
\usepackage{lmodern}
\usepackage[usenames]{color}
\usepackage{subfigure} 

\newcommand{\e}[1]{\mathrm{e}^{#1}}
\newcommand{\vq}{\mathbf{q}}

\newcommand{\cf}{cf. }
\newcommand{\ie}{i.e.}

\begin{document}
\title{Criticality of compact and noncompact quantum dissipative $Z_4$ models in $(1+1)$ dimensions}
\author{Einar B. Stiansen}
\affiliation{Department of Physics, Norwegian University of
Science and Technology, N-7491 Trondheim, Norway}
\author{Iver Bakken Sperstad}
\affiliation{Department of Physics, Norwegian University of
Science and Technology, N-7491 Trondheim, Norway}
\author{Asle Sudb{\o}}
\affiliation{Department of Physics, Norwegian University of
Science and Technology, N-7491 Trondheim, Norway}

\date{Received \today}
\begin{abstract}
\noindent Using large-scale Monte Carlo computations, we study two versions of a $(1+1)$D $Z_4$-symmetric model 
with ohmic bond dissipation. In one of these versions, the variables are restricted to the interval $[0,2\pi\rangle$, 
while the domain is unrestricted in the other version. The compact model features a completely ordered phase with a 
broken $Z_4$ symmetry and a disordered phase, separated by a critical line. The noncompact model features three phases. 
In addition to the two phases exhibited by the compact model, there is also an intermediate phase with isotropic 
quasi-long-range order.
We calculate the dynamical critical exponent $z$ 
along the critical lines of both models to see if the compactness of the variable is relevant to the critical scaling 
between space and imaginary time. There appears to be no difference between the two models in that respect, and we find 
$z\approx1$ for the single phase transition in the compact model as well as for both transitions in the noncompact 
model.
\end{abstract}	
\pacs{75.40.Mg, 64.60.De, 05.30.Rt}

\maketitle

\section{Introduction}
\label{sec:intro}
The standard way of introducing dissipation in a quantum mechanical system is to couple some variable describing the 
system to the degrees of freedom of an external environment\cite{Caldeira-Leggett}. The environment is modeled as a 
bath of harmonic oscillators which couple linearly to the system variables. The oscillator degrees of freedom, appearing 
in the action to second order, may be integrated out to produce an effective theory for the composite system given in 
terms of the system variables. 
 
The presence of a dissipative term introduces strongly retarded (nonlocal in time) self-interactions of the system variables. 
This long-range interaction in imaginary time may have serious consequences for the quantum critical behavior of the system. 
This effect can usually be described by a dynamical critical exponent $z$ defined by the anisotropy of the divergence of the 
correlation lengths at criticality, $\xi_\tau \sim \xi^z$, where $\xi$ and $\xi_\tau$ are the correlation lengths in space and 
imaginary time, respectively. An Ising spin chain with site dissipation was shown by extensive Monte Carlo simulations in 
Ref. \onlinecite{Werner-Volker-Troyer-Chakravarty} to have $z \approx 2$. The same model, augmented to two spatial dimensions, 
was investigated by the present authors in Ref. \onlinecite{PhysRevB.81.104302}. The result $z = 1.97(3)$ suggests that the 
dynamical critical exponent is independent of the number of spatial dimensions, in agreement with naive scaling arguments 
which make no reference to dimensionality.\cite{Hertz_quantum_critical} On the other hand, when coupling the reservoir to 
bond variables involving Ising spins the dissipation term was found to be irrelevant to the universality class, 
\ie, $z \approx 1$.\cite{PhysRevB.81.104302} In general, dissipation suppresses certain types of quantum fluctuations, 
though the larger value of $z$ for site dissipation signifies that bond dissipation is far less effective than site 
dissipation in reducing fluctuations.

Ohmic dissipation in terms of gradients or bonds is common in models describing shunted Josephson junctions 
or granular superconductor systems.\cite{Chakravarty_dissipative_PT,Chakravarty_JJ_array_PRB} Here, the bonds 
represent the difference of the quantum phases between the superconducting grains. In this context it is well-known 
that the coupling of the environment to the system may affect the natural domain of the system variables.\cite{Schön1990237} 
For Josephson junctions, this means that the domain of the phase variables reflects quantization of the charges on each 
superconducting grain. If the charges are quantized in units of Cooper pairs, $2e$, the domain of the quantum phase is 
$2\pi$-periodic. Ohmic shunting leads to an unbounded $-\infty < \theta < \infty$ domain,\cite{Schön1990237} reflecting 
a continuous transfer of charges across the junction. We will henceforth refer to the variable defined on a restricted 
$2\pi$ interval as \emph{compact} and the extended variable as \emph{noncompact}.  

Moreover, dissipation in terms of bonds has also been proposed in an effective model describing the low-energy 
physics of fluctuating loop currents to describe anomalous normal state properties of high-$T_c$ cuprates.
\cite{Aji-Varma_orbital_currents_PRL,Aji-Varma_dissipative_XY} A quantum statistical mechanical model for such degrees 
of freedom has been derived from a microscopic three-band model of the cuprates.\cite{PhysRevB.77.092404} The
classical part of the derived action consists in its original form of two species of Ising variables within each 
unit cell, coupled by a four-spin Ashkin-Teller term. This model has been proven, through large-scale
Monte Carlo simulations, to support a phase transition with a non-divergent non-analyticity in the specific heat 
on top of an innocuous background.\cite{Gronsleth_orbital_currents} The breaking of the Ising-like symmetry 
describes a suggested ordering of loop currents upon entering the pseudogap phase of the cuprates. Neglecting the 
Ashkin-Teller interaction term present in this theory, the classical model may be mapped onto a four-state clock 
model, with the basic variable being an angle parametrizing the four possible current loop
orientations.\cite{Aji-Varma_orbital_currents_PRL,PhysRevB.77.092404,Gronsleth_orbital_currents}

The quantum version of this model includes a kinetic energy term describing the quantum dynamics of the angle 
variables. Adding dissipation of angle differences as in the Caldeira-Leggett approach for Josephson junctions, 
the model has been reported to exhibit local quantum criticality. Local quantum criticality in this context means 
that the model exhibits a fluctuation spectrum which only depends on frequency, but is independent of the
wave vector.\cite{Aji-Varma_orbital_currents_PRL}  This essentially implies a dynamical critical exponent 
$z \rightarrow \infty$. A point which quite possibly is of importance in this context, is that while the starting 
point in Ref. \onlinecite{Aji-Varma_orbital_currents_PRL} is a model with two Ising-like variables, the actual 
dissipative quantum model discussed is one with global $U(1)$ symmetry.  

While the physical picture of fluctuating configurations of current loops suggests an identification of the angles $\theta$ 
and $\theta +2\pi$, the presence of the clearly noncompact dissipation term makes this not entirely obvious. It is {therefore} 
the intent of this work to investigate if the restriction of the variable domain influences the dynamical critical exponent $z$, 
and thereby if it may have consequences for possible manifestations of local quantum criticality in similar models. Since it is 
still an open question exactly what the consequences are of how the variable domains are defined in dissipative quantum models, 
a numerical comparison of the compact and noncompact case is of general interest. We will therefore not restrict the interpretation 
of the model to Ising variables associated with loop currents, although the $Z_4$ symmetry reflecting this starting point will be maintained. Moreover, due to the long-ranged interactions in the imaginary-time 
direction, the Monte Carlo computations are extremely demanding. Since we are interested in a proof of principle of the importance
of compactness versus noncompactness, we will in this paper limit ourselves to a $(1+1)$D model.  

We will perform Monte Carlo simulations on two versions of a dissipative $Z_4$ model described in more detail in Sec. 
\ref{sec:model}, one with compact variables (\ie, a clock model) and one with noncompact variables. The simulation 
details are described in Sec. \ref{Sec:MC}, after which we present the results, first for the noncompact case in 
Sec. \ref{Sec:NC}, then for the compact case in Sec. \ref{Sec:C}.

Our main finding is that, although the critical scaling of space and imaginary time is equal for both cases, \ie, $z = 1$, 
there is a major difference in phase structure. Whereas the compact model displays a conventional order-disorder phase 
transition, the noncompact model develops an intermediate phase characterized by power-law decay in spin correlations 
(quasi-long-range order) and a $U(1)$ symmetric {distribution} of the complex order parameter. The appearance of this 
intermediate phase is related to the fact that the kinetic energy term must be treated differently for the compact and 
noncompact cases, as we discuss in detail in the Appendix.

It is well established that this kind of critical phase occurs in classical 2D $Z_q$ clock models and $XY$ models with $Z_q$ 
anisotropy,\cite{Elitzur_clock_model,Jose-Kadanoff-Kirkpatrick-Nelson_XY,PhysRevB.23.1357} but only for larger {values of 
$q$} than we are considering. It is remarkable that the noncompact model presented in this paper exhibits a critical phase 
with emergent $U(1)$ symmetry, when the dissipationless starting point is a pure $Z_4 = Z_2 \times Z_2$ model (\ie, a double 
Ising model) with the angle variables restricted to four discrete values by a \emph{hard} constraint. We will discuss this 
in more detail in Sec. \ref{Sec:discussion}, after which we summarize our results in Sec. \ref{Sec:concl}.

\section{The model}
\label{sec:model}

The starting point for our model is a chain of $N_x$ quantum rotors, or equivalently planar spins, the alignment of 
which is described by a set of angle variables $\lbrace \theta_x \rbrace$. Although these variables could also be 
denoted as the phases of the quantum rotors, we will refer to them simply as angles. Requiring that the spins 
satisfy $Z_4$ symmetry, the angles can be parametrized as $\theta = 2\pi n /4$ with integer $n$, making our model 
similar to a four-state (or $Z_4$) clock model. Being quantum spins, their dynamics is described by their evolution 
in imaginary time $\tau$, with $N_\tau$ denoting the number of Trotter slices used to discretize the imaginary 
time dimension. The variables $\lbrace \theta_{x,\tau} \rbrace$ are thus defined on the vertices of a $(1+1)$D 
quadratic lattice of size $N_x \times N_\tau$.

In order to investigate if the restriction on the angle variable is relevant to the dynamical critical exponent 
$z$ or not, we will consider two variants of this model, with the complete action for both stated below for later 
reference. In the compact (C) case, we restrict the parametrization variable $n$ to just four values, so that the 
angle $\theta$ is restricted to one primary interval, corresponding to the four primary states of the four-state 
clock model. In the noncompact (NC) case we have no such restriction, and $n$ can take any integer values. The 
general form of the action is 
\begin{align}
S^\mathrm{C,NC} = S_{\tau}^\mathrm{C,NC} + S_{x} + S_{\mathrm{diss}},
\end{align}
where the kinetic energy for the compact and the noncompact case, respectively, is given by
\begin{align}\label{term_SC}
S_{\tau}^\mathrm{C} &= -K_\tau \sum_{x=1}^{N_x}\sum_{\tau=1}^{N_\tau}\cos(\theta_{x,\tau+1}- \theta_{x,\tau}),
\end{align}
\begin{align}\label{term_SNC}
S_{\tau}^\mathrm{NC} &= \frac{K_\tau}{2} \sum_{x=1}^{N_x}\sum_{\tau=1}^{N_\tau}(\theta_{x,\tau+1}- \theta_{x,\tau})^2.
\end{align}
{The spatial interaction defines a periodic potential}
\begin{align}\label{SX} 
S_x=-K\sum_{x=1}^{N_x}\sum_{\tau=1}^{N_\tau}\cos(\theta_{x+1,\tau}- \theta_{x,\tau}),
\end{align}
{and the dissipation term is defined according to}
\begin{align}\label{SDISS}
S_{\mathrm{diss}} = \frac{\alpha}{2}\sum_{x=1}^{N_x}\sum_{\tau\neq\tau'}^{N_\tau}\left(\frac{\pi}{N_\tau}\right)^2
\frac{\left(\Delta\theta_{x,\tau}   -  \Delta\theta_{x,\tau'}\right)^2}
{\sin^2(\frac{\pi}{N_\tau}|\tau-\tau'|)}.
\end{align}
The bond variable or angle difference is written as $\Delta\theta_{x,\tau} = \theta_{x+1,\tau} - \theta_{x,\tau}$.

Note that the only apparent difference between the compact and the noncompact model is the form of the kinetic 
energy term. When the angles are compact the short-range temporal interaction is given by a cosine term, 
in contrast to noncompact angles for which a quadratic form of the kinetic term must be used. 
The reason for this difference can be traced to the fact that, whereas canonical conjugate variables of compact 
angles are discrete due to the $2\pi$ periodicity of the quantum wave functions, no such restriction applies when 
the angles are noncompact. From a qualitative point of view the two separate forms of the temporal interaction 
term is expected. Considering the imaginary time history of a single variable, it is clear that a cosine interaction 
in imaginary time will render the ground state of the noncompact model massively degenerate. A Trotter slice 
may be shifted by $2\pi$ relative to the neighboring Trotter slices without any penalty in the action. However, 
a quadratic interaction term in the imaginary time direction lifts this degeneracy and tends to localize the 
angle variables.

There is nothing new about the derivation of these different kinetic terms, but as the difference is crucial to the 
phase structure of our models and is also rarely discussed in the literature, we include the derivation in the 
Appendix. In addition, in order to simulate the compact model we also need an appropriate reinterpretation of 
the dissipation term. We find it natural to postpone this to Sec. \ref{Sec:C}.

The action is on a form identical to the model in Sec. III in Ref. \onlinecite{PhysRevB.81.104302} apart from the 
nature of the variables and the resultant treatment of the dissipation term. However, we still expect the scaling 
arguments presented in  Ref. \onlinecite{PhysRevB.81.104302} to be valid since no reference to the actual type of 
variable is used. The action in Fourier space {may be written}
\begin{align}
	S \sim ( \vq^2 + \omega^2 + |\omega|\vq^2) \theta_{q} \theta_{-q},
\end{align}
neglecting any prefactors. Taking the limit $\vq \rightarrow 0 $, $\omega \rightarrow 0$ we anticipate that the term 
$\sim |\omega|\vq^2$ describing the dissipation is subdominant for all positive $z$. Accordingly, we expect at least 
naively that $z=1$ for both compact and noncompact variables. This will be investigated in detail in 
our simulations, and we make no assumption of the veracity of naive scaling applied to this problem.

\section{Details of the Monte Carlo computations}
\label{Sec:MC}

When expanding the dissipative term, it becomes clear that it contributes both to ferromagnetic and antiferromagnetic long 
range interactions. This renders the system intractable to the Luijten-Blöte\cite{Luijten-Blote} extension of the Wolff 
cluster algorithm\cite{Wolff} which has been used with great success in systems with noncompeting interactions. Also, for 
the case of noncompact variables there does not exist a straightforward way of defining (pseudo)spin projections, a necessary point 
for the Wolff embedding technique.\cite{Wolff} Considerable progress has been made in constructing new effective algorithms for long 
range interacting systems with extended variables.\cite{PhysRevLett.95.060201, 1742-5468-2005-12-P12003, Werner_dissipative_MC_algorithms} 
However, these algorithms are presently restricted to $(0+1)$D systems, and do not seem to generalize easily to 
$N_x > 1$.\cite{1742-5468-2005-12-P12003} Furthermore, the basic degrees of freedom in these algorithms are the phase 
\emph{differences} between two superconducting grains in an array of Josephson junctions. Our aim is to investigate the 
ordering of the phases themselves. Hence, the existing non-local algorithms may not be utilized. In the Monte Carlo 
simulations, we have therefore used a parallel tempering algorithm\cite{Hukushima_parallel_tempering,Katzgraber_MC} 
in which several systems (typically 16 or 32) are simulated simultaneously at different coupling strengths. 

A Monte Carlo sweep corresponds to proposing a local update by the Metropolis-Hastings algorithm  for every grid point in 
the system in a sequential way. For the case of noncompact variables the proposed new angles are 
generated by randomly choosing 
to increase or decrease the value, then propagating the value by randomly choosing the increment on the interval
$\{\frac{\pi}{2},\pi,\frac{3\pi}{2},2\pi \}$. In the case of compact variables, a new angle value in the primary interval is 
randomly chosen. After a fixed number of Monte Carlo sweeps (typically $3 - 10$) a parallel tempering move is made. In this move, a swap of 
configurations between two neighboring coupling values is proposed, and the swap is accepted with probability $\Xi_{PT}$ 
given by 
\begin{equation}
 \Xi_{PT} =
\begin{cases}
1 &\mathrm{if} \ \Delta < 0, \\
\e{-\Delta} &\mathrm{if} \ \Delta \geq 0.
\end{cases}
\end{equation}
Here, $\Delta = \kappa'(\bar{S}[X;\kappa']- \bar{S}[X';\kappa']) - \kappa(\bar{S}[X;\kappa] - \bar{S}[X';\kappa]) $, where $\kappa$ 
is the coupling value varied, representing in our case $K$ or $\alpha$, and $X$ represents the angle configuration. $\bar{S}$ 
indicates the term of the action proportional with the coupling parameter $\kappa$.

All Monte Carlo simulations were initiated with a random configuration. Depending on system sizes various numbers of sweeps were performed for each coupling value. For the phase transition separating the disordered state from the critical phase in the noncompact model $5-10\times 10^6$ sweeps were made. Also, $1-5\times10^5$ sweeps at each coupling 
value were discarded for equilibration. For the compact model and the second transition of the noncompact model as much as $30\times10^6$ sweeps where performed and typically $5\times10^5$ sweeps discarded.

 The Mersenne-Twister\cite{Mersenne_Twister} random number generator was used in all 
simulations and the random number generator on each CPU was independently seeded. It was confirmed that other random number generators yielded consistent results. We also make 
use of the Ferrenberg-Swendsen reweighting technique,\cite{PhysRevLett.63.1195} which enables us to continuously vary the coupling 
parameter after the simulations have been performed.
 
\section{Results: Noncompact model}
\label{Sec:NC}

In this section we consider the noncompact version of the dissipative $Z_4$ model. Using 
Eqs. \eqref{term_SNC}, \eqref{SX} and \eqref{SDISS} we have the following action,
\begin{align} \label{SNC} 
S^\mathrm{NC} = S_\tau^\mathrm{NC} + S_x +S_{\mathrm{diss}}.
\end{align}
In contrast to the compact model, the angle variables are in this case not restricted to the primary interval. The variables 
are straightforwardly generalized to take the values $\theta = 2\pi n/4$, where $n = 0,\pm 1, \pm 2, \dots$. We seek to fix 
$K$ and $K_\tau$ and investigate how the system behaves under the influence of increasing dissipation strength controlled 
by the dimensionless parameter $\alpha$. 

\begin{figure}
  \centering
   \includegraphics[width=0.45\textwidth]{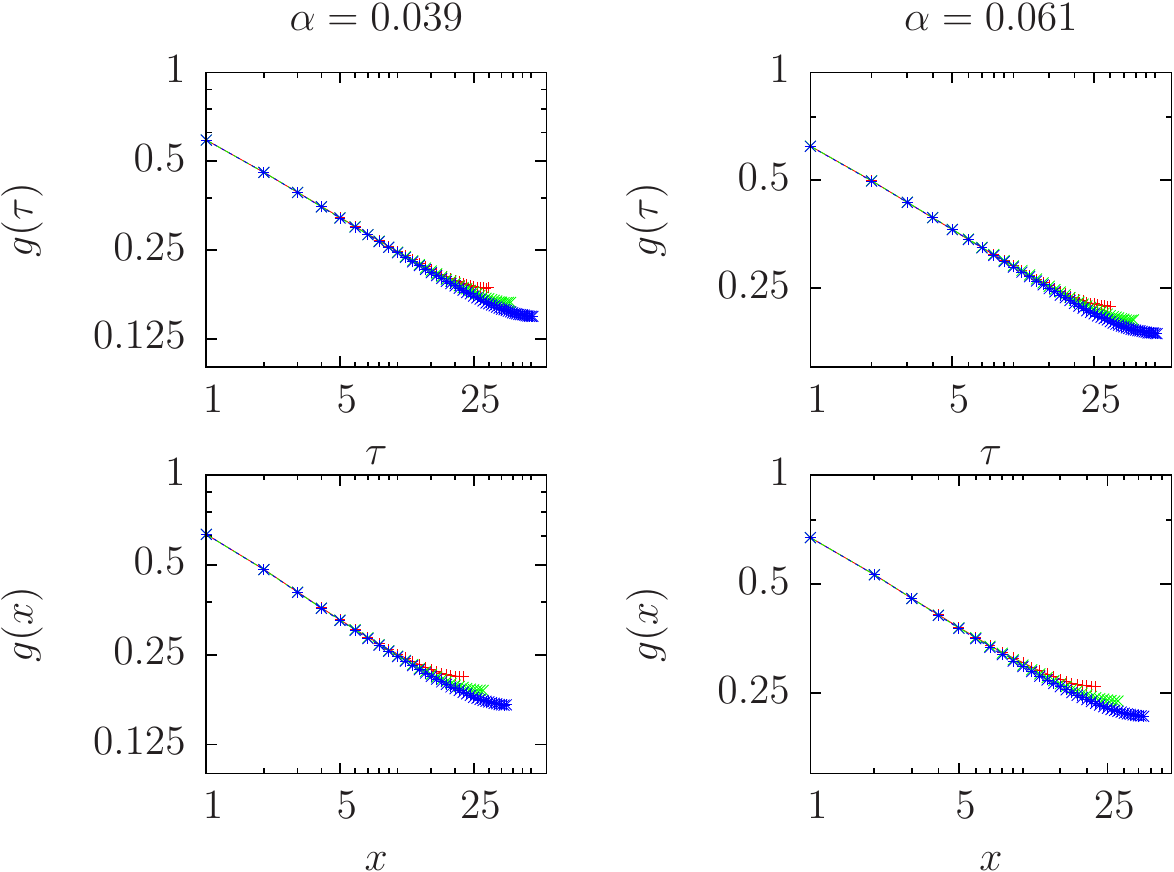}
  \caption{(Color online) Correlation functions, Eq. \eqref{spin_corr}, at two
values of the dissipation strength $\alpha$ within the critical phase
for spatial coupling $K=0.75$. System sizes are $N_x = 44,57,74$, with
optimal choices of $N_\tau$ at $\alpha_c^{(1)}$, see text. \emph{Top
row:} Correlation functions for the temporal direction. \emph{Bottom
row:} Correlation functions for the spatial direction. } 
  \label{corr_table}
\end{figure}

The kinetic coupling strength has been fixed to $K_\tau =0.4$ for computational reasons, as this ensures that the simulations 
will be performed at convenient values of $N_x$ and $N_\tau$. We have performed simulations at four different spatial coupling 
constants $K = $ 0.4, 0.5, 0.6 and 0.75. These choices are also made for computational convenience, as the limit of 
vanishing dissipation as well as the limit $K \to 0$ are both very computationally demanding. For all coupling values 
there is a disordered phase at low values of the dissipation strength. In this phase the noncompact angles exhibit wild 
fluctuations and consequently  
$\langle \e{\mathrm{i}\theta_{x,\tau}}\rangle = 0$. However, we also have $\langle \e{\mathrm{i}\Delta\theta_{x,\tau}}\rangle \neq 0$ 
in this phase, a trivial consequence of the cosine potential acting as an external field on the bond variables. The bond variables
occasionally drift from one minimum of the extended cosine potential to another. As the dissipation strength is 
increased, fluctuations in these variables are suppressed, and the system features two consecutive phase transitions separated 
by a critical phase. This intermediate phase is characterized by power-law decay of spatiotemporal spin correlations on the form 
\begin{align}\label{spin_corr}
g(\mu) = \langle \e{\mathrm{i}(\theta_\mu -\theta_0)}\rangle, \hspace{3mm}  \mu \in (x,\tau). 
\end{align}
The correlation functions for both spatial and imaginary time direction are shown in Fig. \ref{corr_table} for two different dissipation strengths both within the the critical phase.

A very similar critical 
phase, as well as phase transitions associated with it, has recently enjoyed increased interest in various versions of 
classical clock models.\cite{PhysRevLett.96.140603,Baek-Minnhagen_5-state,Baek-Minnhagen-Kim_8-state} We will proceed under 
the assumption that a similar picture is valid in our case. Indeed, simulations performed on a classical 2D six-state clock 
give qualitatively very similar results for all observables considered below, which supports the {supposition} that these 
two phenomena are related.

\begin{figure}
    \centering
    \subfigure[Two dimensional Gaussian distribution of the order parameter in the complex plane corresponding 
    to the disordered phase with $\alpha=0.0<\alpha_c^{(1)}$.]
    {
        \includegraphics[width=0.4\textwidth]{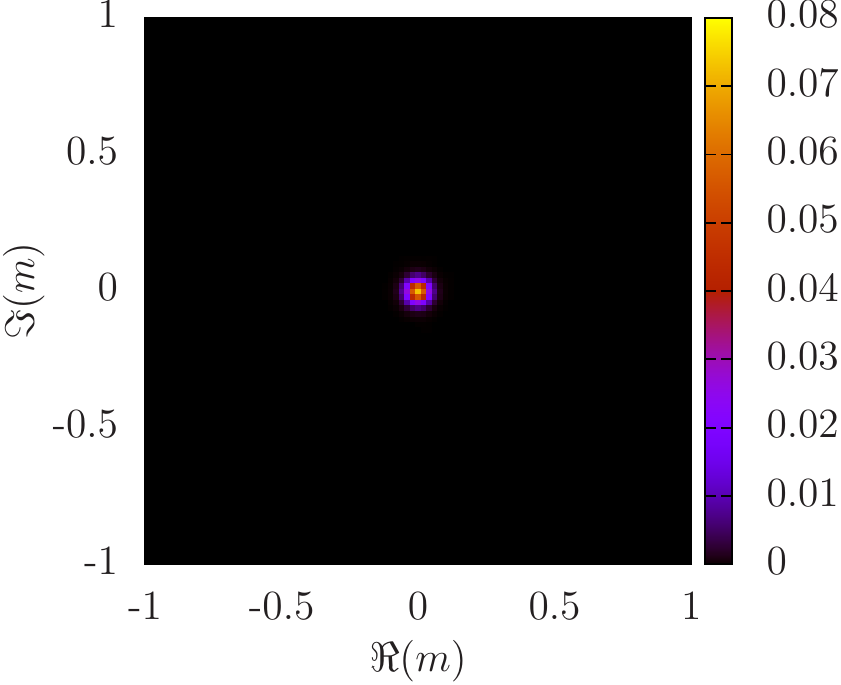}
        \label{fig:gaussian}
    }
    \\
    \subfigure[Intermediate critical phase exhibiting a finite-size-induced non-vanishing $|m|$ that rotates in the $\phi$ 
    direction. The critical phase exists in a finite interval of dissipation strengths 
    $\alpha_c^{(1)}<\alpha=0.04<\alpha_c^{(2)}$. The remaining anisotropy is attributed to insufficient sampling.
    \cite{Hove_clock_model}]
    {
        \includegraphics[width=0.4\textwidth]{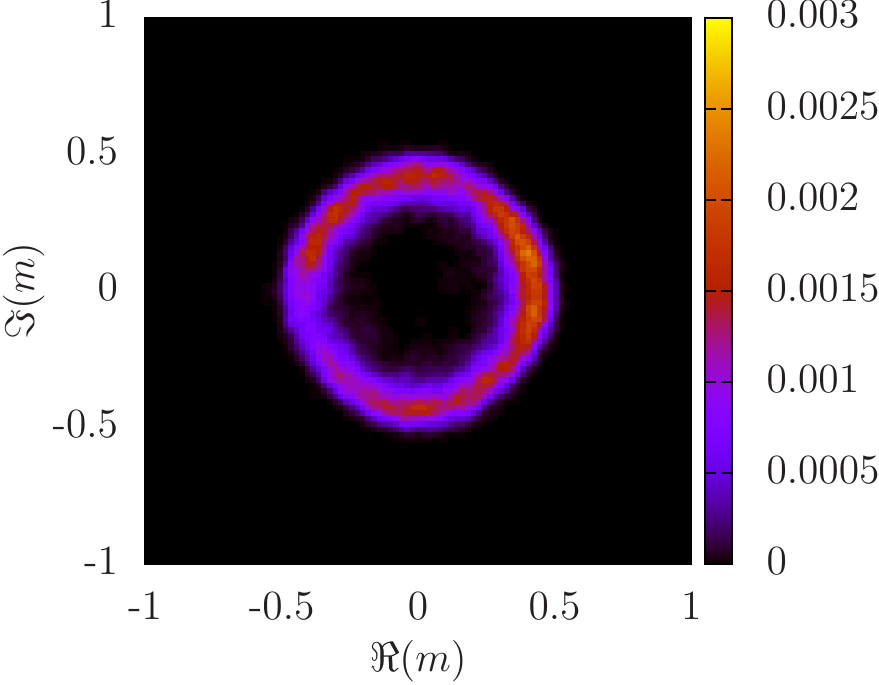}
        \label{fig:quasi}
    }
    \subfigure[The rotational symmetry of the intermediate critical phase is broken and long-range order is established as the 
    order parameter relaxes into {one of the four directions in the complex plane}. The long-range ordered phase corresponds 
    to the strong dissipation limit, $\alpha = 0.18>\alpha_c^{(2)}$.]
    {
        \includegraphics[width=0.4\textwidth]{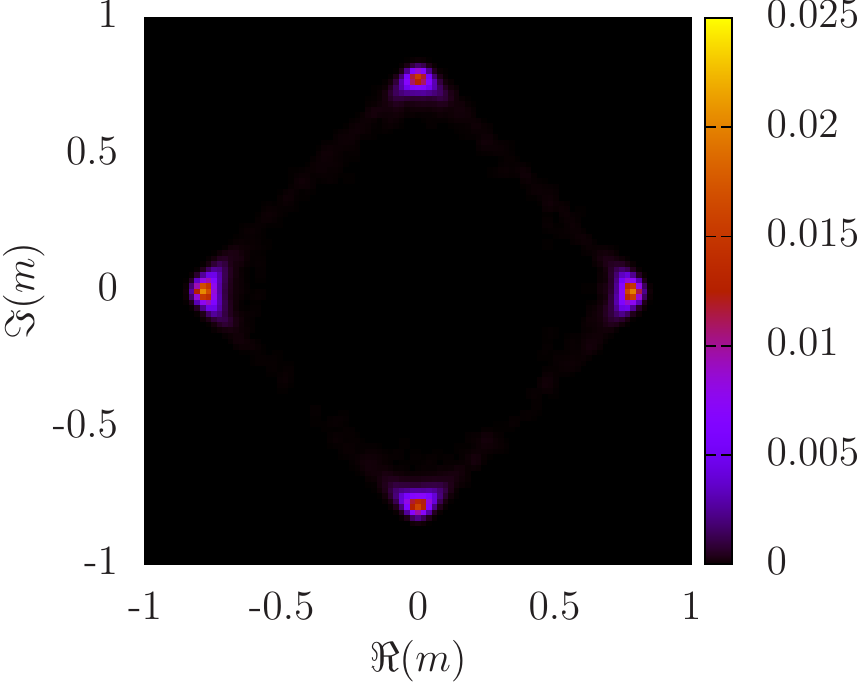}
        \label{fig:order}
    }
    \caption{(Color online) Evolution of the complex order parameter when dissipation strength $\alpha$ is increased for 
    $K=0.75$ and system size $N_x=74$, $N_\tau=103$ which corresponds to a near optimal aspect ratio at the phase 
    transition at $\alpha\approx\alpha_c^{(1)}$. The color scale indicates relative density of the distribution.}
    \label{fig:sample_subfigures}
\end{figure}

Considering the complex order parameter of the system,
\begin{align} \label{order_parameter}
m = \frac{1}{N_x N_\tau}\sum_{x,\tau}\e{\mathrm{i}\theta_{x,\tau}} = |m|\e{\mathrm{i}\phi},
\end{align}
the intermediate critical phase can be identified by observing the distribution of $m$ in the complex plane.
\cite{Baek-Minnhagen-Kim_8-state} In the disordered 
phase, the order parameter is a Gaussian peak centered at the origin. In the intermediate phase, quasi-long-range order develops in 
the complex order parameter, and so $|m|$ acquires a nonzero value as a finite-size effect. 
The order parameter is, however, free to rotate in the $\phi$ direction. This can be described as the vanishing
of the excitation gap naively expected for discrete $Z_q$ models, or equivalently as an emergent $U(1)$ 
symmetry.\cite{Lou-Sandvik-Balents_anisotropy} 
{This symmetry is broken at a larger value of the dissipation strength, when true long-range order is established when 
the magnetization selects one of the four well-defined directions in the complex plane originating with the underlying 
$Z_4$ symmetry.} Typical distributions of the complex order parameter in the three phases is shown in Fig. \ref{fig:sample_subfigures}.

\begin{figure}
  \centering
   \includegraphics[width=0.45\textwidth]{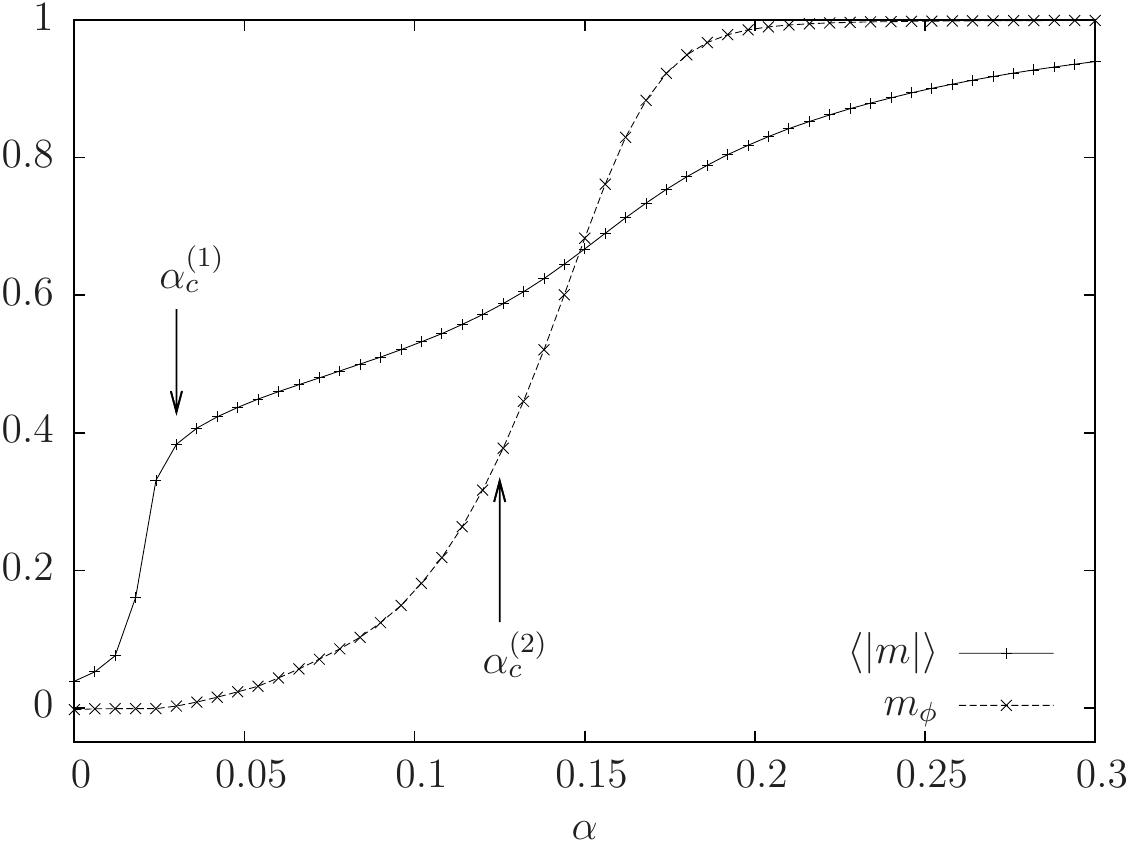}
  \caption{{The order parameter $\langle |m| \rangle$ and the anisotropy measure $m_\phi $ of the noncompact 
  $Z_4$ model 
  with $K=0.75$ and system size $N_x=74$, $N_\tau=103 \approx N_\tau^\ast$.  This size represents a near optimal aspect 
  ratio at $\alpha\approx\alpha_c^{(1)}$.} The two phase transitions are indicated by arrows, note that the intermediate 
  critical phase $\alpha_c^{(1)}<\alpha<\alpha_c^{(2)}$ features a rotationally symmetric order parameter
  distribution.}
  \label{order_param}
\end{figure}

Although not presented here, we have also confirmed that the susceptibility of the order parameter diverges over a 
finite interval of dissipation strengths, also a clear evidence of a critical phase. 

The phase transition between the disordered state and the intermediate critical phase at dissipation strength 
$\alpha=\alpha_c^{(1)}$ is detected by the Binder cumulant $g=1-Q/3$, where 
\begin{align}\label{Binder}
 Q=\frac{\langle |m|^4  \rangle}{\langle |m|^2 \rangle^2}.
\end{align}
The brackets indicate ensemble averaging. The scaling at criticality of the Binder cumulant for anisotropic 
systems is given in terms of two independent scaling variables,\cite{Werner-Volker-Troyer-Chakravarty}
\begin{align}\label{scalebinder}
g(N_x,N_\tau) = \mathcal{G}\left(\frac{N_x}{\xi},\frac{N_\tau}{\xi^z}\right).
\end{align} 
At a critical point the correlation length $\xi$ diverges, and one should be able to observe data collapse of the Binder 
cumulant as a function of $N_\tau/N_x^z$ for the correct value of $z$.  The value of $g(N_x,N_\tau)$ is independent of 
$N_x$ at the critical coupling, this may be used to align the plots of $g$ as a function of $N_\tau$ horizontally. 
The exponent $z$ can then be found by optimal 
collapse of data onto a universal curve. The cumulant curves have a maximum at $N_\tau = N_\tau^\ast$. At this temporal size, 
the system appears as isotropic as it can be, the anisotropic interactions taken into account. See 
Ref. \onlinecite{PhysRevB.81.104302} for a thorough discussion of this finite-size analysis.

In the intermediate phase, the system is critical over a finite interval 
of dissipation strengths. According to the scaling Eq. \eqref{scalebinder}, curves of the Binder cumulant for increasing 
system sizes will therefore merge in this interval for $N_x \rightarrow \infty$.\cite{Loison_KT_Binder} 
For systems of finite sizes as considered here, the curves will however intersect close to the transition instead,
and we find $\alpha_c^{(1)}$ by inspecting the convergence of the crossing points. As discussed in Ref. ~\onlinecite{PhysRevB.81.104302}, the functional form of this convergence is unknown in our case (\cf also Sec. \ref{Sec:discussion} and Ref. ~\onlinecite{Loison_KT_Binder}), and all we can do is to report our best estimate for the $N_x \rightarrow \infty$ transition point. The uncertainty estimated accordingly is not insignificant, but the effective critical exponent $z$ is found to not be very sensitive to this error in $\alpha_c$. 

By further increasing the dissipation strength, the rotational symmetry of the global order parameter is broken at 
$\alpha=\alpha_c^{(2)}$. The Binder cumulant given by Eq. \eqref{Binder} will not pick up this transition because 
$|m|$ does not contain any information on the angular direction of the global magnetization. Therefore, we 
consider an alternative magnetization measure\cite{Lou-Sandvik-Balents_anisotropy, Baek-Minnhagen-Kim_8-state} 
\begin{align}\label{disc_mag}
m_\phi = \langle \cos(4\phi) \rangle,
\end{align}
where $\phi$ is the global phase as indicated by Eq. (\ref{order_parameter}). This anisotropy measure vanishes when $\phi$ is evenly 
distributed and tends toward unity when the excitation gap opens and $\phi$ gets localized. 
We show in Fig. \ref{order_param} both order parameters 
for the system $N_x=74,N_\tau=103$ as a function of $\alpha$. This $N_\tau$ corresponds to the nearest 
integer $N_\tau^\ast$ at $\alpha\approx\alpha_c^{(1)}$. Actually, the optimal $N_\tau$ decreases with increasing $\alpha$, 
so the given system size does not represent an optimally chosen aspect ratio for other dissipation strengths. The rotational 
symmetry of the complex order parameter is clearly seen to be broken at a higher dissipation strength than the onset of the 
intermediate critical phase.

\begin{figure}
  \centering
   \includegraphics[width=0.45\textwidth]{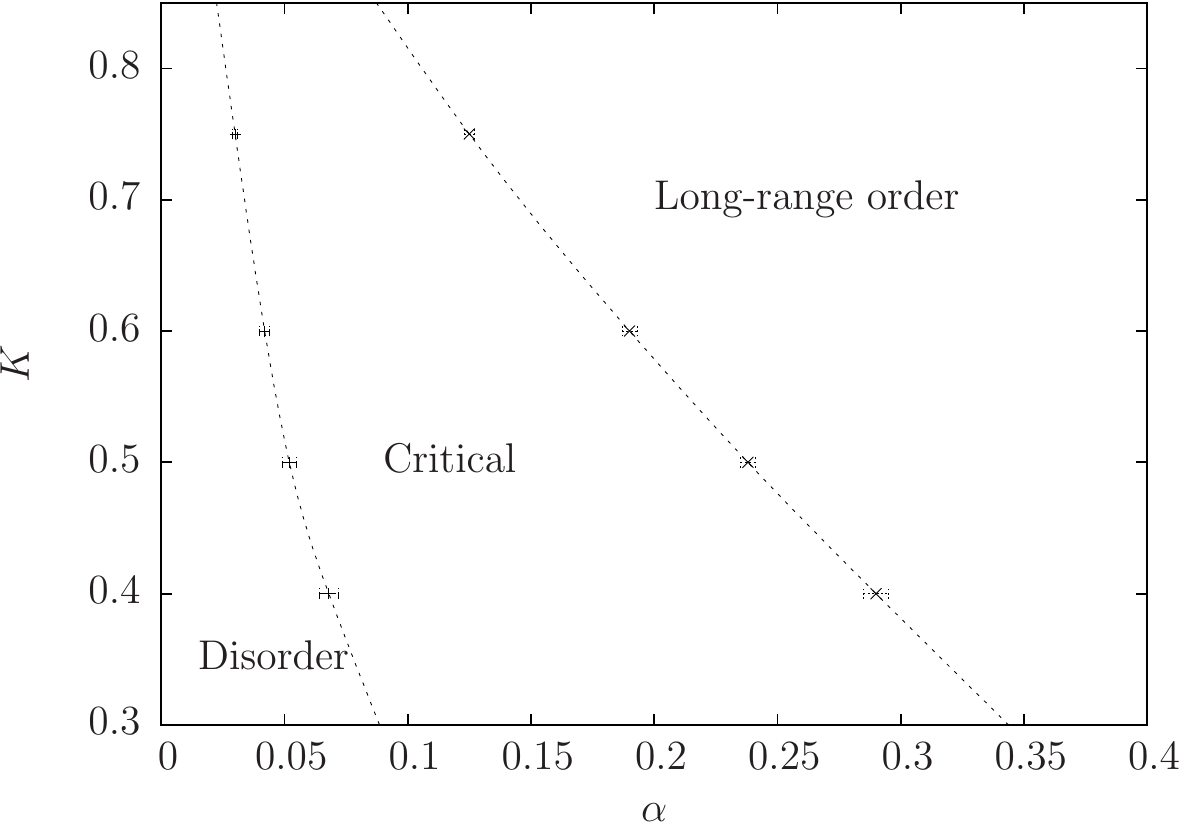}
  \caption{Phase diagram for the noncompact model, Eq. \eqref{SNC} with $K_\tau = 0.4$. The dotted 
  lines are guides to the eye. For fixed $K$ the model 
  features two consecutive phase transitions surrounding the intermediate critical phase 
  (with quasi-long-range order). The simulation results (symbols along the dotted lines) 
  are restricted to a region in coupling space amenable to simulations. }
  \label{diagramNC}
\end{figure}

Because $\phi$ measures a global rotation of the order parameter, extremely long simulations is needed to explore the $\phi$ space 
with a local update algorithm. This limits the efficiency of constructing a Binder cumulant from $m_\phi$ and extracting $\alpha_c^{(2)}$ 
from a universal point because this would involve calculating moments of a already statistically compromised ensemble. To alleviate these difficulties, we instead make 
a scaling ansatz for the anisotropy measure itself,
\begin{align}\label{disc_Mag}
m_\phi= \mathcal{M}_\phi \left(\frac{N_x}{\xi},\frac{N_\tau}{\xi^z}\right),
\end{align}
based on the fact that the naive scaling dimension of this magnetization measure is zero. 
Near criticality, we expect $m_\phi$ to scale with system size in the same way as the Binder cumulant Eq. \eqref{scalebinder}. 
Hence, we may calculate a dynamical critical exponent for this transition by exactly the same procedure as in Sec. \ref{Sec:C} 
and Ref. \onlinecite{PhysRevB.81.104302}. Again we expect a merging of $m_\phi$ curves  as $\alpha \rightarrow \alpha_c^{(2)}$ 
from above in the limit of large $N_x$, but for the present system sizes we use the crossing points of $m_\phi$ curves to 
estimate $\alpha_c^{(2)}$. In Fig. \ref{diagramNC}, we plot the resulting phase diagram in the $\alpha-K$ plane.
The intermediate phase is evidently very wide also when compared to the uncertainty assigned to 
the transition line, and we feel confident that it is a genuine phase and not merely an effect of the admittedly moderate
finite system sizes we are restricted to.
  
\begin{figure}
  \centering
   \includegraphics[width=0.45\textwidth]{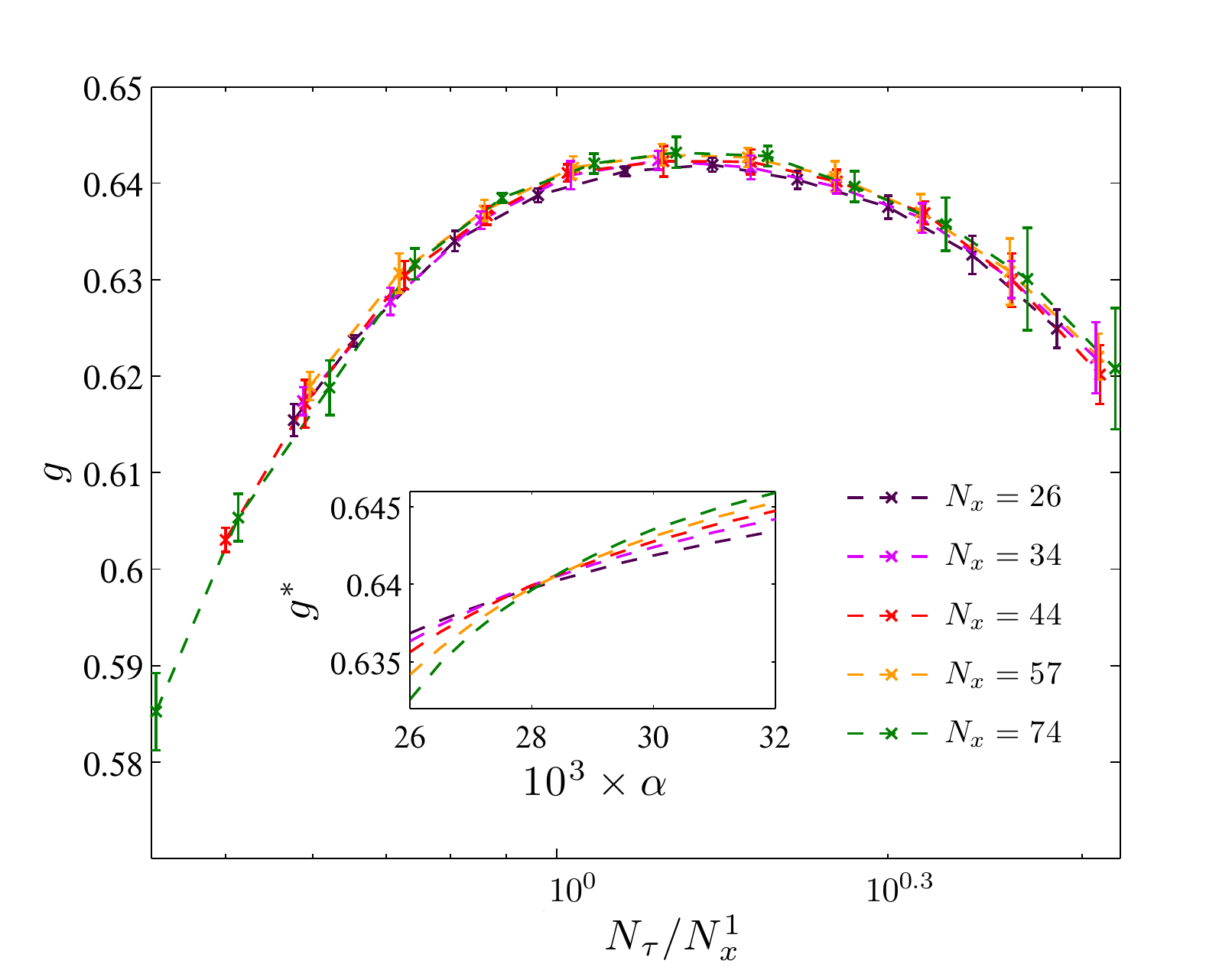}
  \caption{(Color online) Data collapse of the Binder cumulant, $g=1-Q/3$, with $Q$ given by Eq. (\ref{Binder}), for the 
   noncompact $Z_4$ model at $K=0.75$ and $\alpha =0.030 \approx \alpha^{(1)}_c $ with $z^{(1)}=1$. Inset: Intersection of the Binder cumulant as a function of dissipation strength.}
  \label{binder075}
\end{figure}

\begin{figure}
  \centering
   \includegraphics[width=0.45\textwidth]{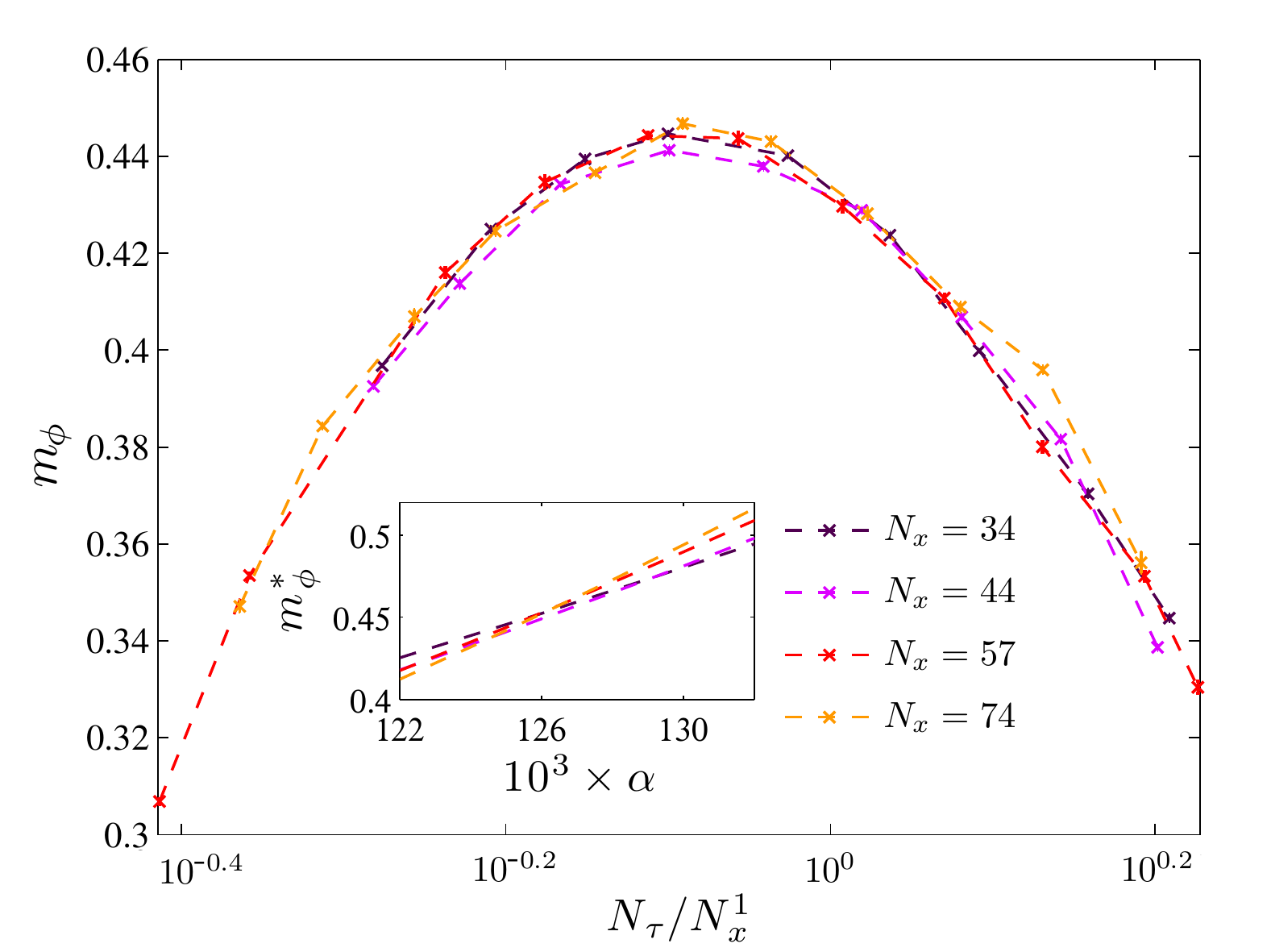}
  \caption{(Color online) Data collapse of the anisotropy measure $m_\phi$, Eq. \eqref{disc_mag}, for the noncompact $Z_4$ model 
  at $K=0.75$ and $\alpha =0.125 \approx \alpha_c^{(2)}$ with $z^{(2)}=1$. The actual uncertainties are probably larger than indicated by the error bars for reasons discussed in the text. Inset: Intersection of the anisotropy measure as a function of dissipation strength.}
  \label{Mag075}
\end{figure}

We extract the dynamical critical exponent $z$ along both of the critical lines $\alpha_c^{(1)}$ and $\alpha_c^{(2)}$ 
for all spatial coupling strengths. The data collapse of the Binder cumulant $g$ at $K=0.75$ and 
$\alpha = 0.030\approx\alpha_c^{(1)}$ is shown in Fig. \ref{binder075}. Increasing the dissipation strength further 
brings the system to the second phase transition at $\alpha = 0.125 \approx \alpha_c^{(2)}$, the collapse of $m_\phi$ 
at this point is shown in Fig. \ref{Mag075}.
 
In Table \ref{tab:table}, we present the numerical estimates of the dynamical critical exponent. The values of 
$z$ are obtained using the scaling relation $N_\tau^\ast = a N_x^z$, with uncertainties based on a bootstrap 
analysis. 
These uncertainties also include the uncertainty in $\alpha_c$.
Within the accuracy of the simulations, the value of the critical exponent 
is $z=1$ for all the coupling values at both phase transitions (although precise results are harder to obtain 
for the second). This is in accordance with the scaling argument presented in Sec. \ref{sec:model}.

\begin{table}
\caption{Numerical estimates for critical coupling and critical
exponents $z^{(1),(2)}$ for the two phase
transitions $\alpha_c^{(1),(2)}$ of the noncompact model.}
\begin{center}
    \begin{tabular*}{0.5\textwidth}{@{\extracolsep{\fill}}  c  c  c  c 
c}
    \hline
    \hline
    $K$ & $\alpha_c^{(1)}$ & $z^{(1)}$  & $\alpha_c^{(2)}$ & $z^{(2)}$
\\ \hline
    0.75 & 0.030(2) & 0.99(1) & 0.125(2) & 1.01(2) \\
    0.6  &  0.042(2)& 0.99(2) & 0.190(3) & 0.96(3) \\
    0.5  & 0.053(2) & 1.02(2) & 0.238(3) & 0.97(3) \\
    0.4  & 0.068(4) & 0.97(3) & 0.287(5) & 0.99(4) \\
    \hline
    \hline
    \end{tabular*}\label{tab:table}
\end{center}
\end{table}

\section{Results: Compact model}
\label{Sec:C}

We now turn to the compact version of the dissipative $Z_4$ model,
\begin{align}
	\label{SC}
	S^\mathrm{C} = S_\tau^\mathrm{C} + S_x + S_{\mathrm{diss}},
\end{align}
where the three terms are given by Eqs. \eqref{term_SC}, \eqref{SX} and \eqref{SDISS}, respectively. 
Note that we now use a kinetic term $S_\tau^\mathrm{C}$ having the same cosine-form as the spatial interaction term 
$S_x$. Regarding the use of the same dissipation term $S_{\mathrm{diss}}$ as in the noncompact case, one may argue  
that adding a Caldeira-Leggett term for the angle differences $\Delta \theta$ is a rather artificial way to model
dissipation for a compact clock model in the first place, since its variance under $2\pi$ translations of 
$\theta$ implicitly assumes noncompact variables. However, adding exactly such a dissipation term is crucial 
for the demonstration of local quantum criticality in a similar $Z_4$ model\cite{Aji-Varma_orbital_currents_PRL} 
that is not obviously noncompact. Therefore, our motivation for the comparative study in the present section of a 
compactified version of the action \eqref{SNC} is to investigate whether an equivalent dissipation term for compact 
variables gives the model the same critical properties as reported for noncompact variables in the previous section, 
and thus whether the compactness of the variables as such is essential. Constructing an appropriate compactified 
version of the dissipative model does, however, require a reinterpretation of the variables in the Caldeira-Leggett 
term, so we will begin with a careful discussion of how we should treat this term in our simulations.

We first impose the following restriction on the interpretation of the compactified dissipation term: The term as 
a whole should be invariant under translations $\theta \rightarrow \theta + 2\pi$, since these two states are 
indistinguishable. As a corollary, any configurations that are physically indistinguishable when the angles are 
restricted to four values $\theta \in \lbrace -\pi, -\pi/2, 0, \pi/2 \rbrace$ (or any equivalent parametrization) 
should give the same contribution to the dissipation term. Consequently, we cannot simply simulate the model 
with the dissipation term \eqref{SDISS} as it stands, because the angle differences $\Delta \theta_{x,\tau}$ 
now only make physical sense modulo $2\pi$. We therefore have to bring $\Delta \theta_{x,\tau}$ back to the 
primary interval $\lbrack -\pi, \pi \rangle$, as is well known for phase differences in superconducting systems without 
dissipation and other realizations of the (compact) $XY$ model. Furthermore, we also choose to do the same for 
the difference between the two (compactified) $\Delta \theta_{x,\tau}$ terms in Eq. \eqref{SDISS}, as the 
alternative would result in different Boltzmann factors being associated with physically equivalent situations.
Our procedure then is equivalent to requiring that the entire difference 
$\Delta \theta_{x,\tau} - \Delta \theta_{x,\tau'}$ should be restricted to the primary interval $\lbrack -\pi, \pi\rangle$, 
\ie, treating the dissipation term as a $2\pi$-periodic function. 

The details of the Monte Carlo simulations are described in \ref{Sec:MC} also for the compact model. The only 
difference that may be of any consequence is that we found it more convenient to vary the spatial coupling 
while fixing the dissipation strength in this case, but we have checked that the direction in coupling space 
taken by the simulations has no impact on the result. 

The dissipationless ($\alpha=0$) four-state clock model is completely isomorphic to the Ising model with interaction 
$K/2$. Thus, we may employ the criterion $\sinh(K_c)\sinh(K_\tau)=1$ in order to calculate $K_c$ for a fixed value of 
$K_\tau$. The temporal coupling parameter is fixed at $K_\tau = -\ln{(\tanh{\frac{1}{2}})} \approx 0.7719$ such that 
$K_c = 1$ when the dissipation is tuned to zero. 

The most striking difference we found when compactifying the angles is that the intermediate phase with 
quasi-long-range order vanishes. This means that one has only a single disorder-order phase transition, as 
is the result one would usually expect for any model with $Z_4$ symmetry. We have verified that the $Z_4$ symmetry 
and the apparent $U(1)$ symmetry of the complex order parameter (in the disordered phase) are spontaneously 
broken simultaneously at a single critical point. This is found by observing that the inflection points of 
magnetization curves for $m$ and $m_\phi$ coincide asymptotically, in contrast to the curves shown in 
Fig. \ref{order_param} for the noncompact case.

\begin{figure}
  \centering
   \includegraphics[width=0.45\textwidth]{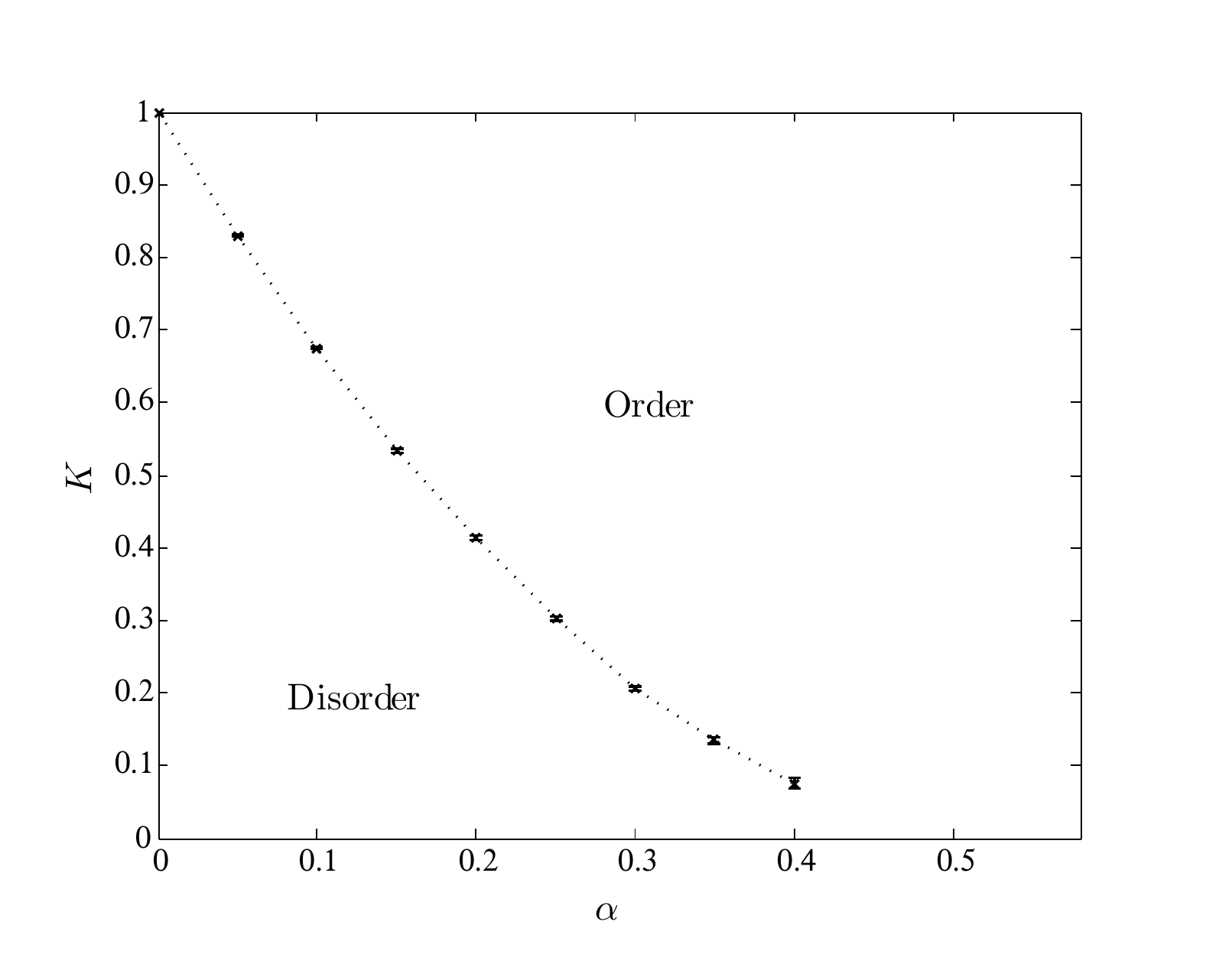}
  \caption{Phase diagram for the compact model of Eq. \eqref{SC} with $K_\tau = -\ln{(\tanh{\frac{1}{2}})}$, 
   the dotted line indicating a critical line separating the disordered phase from a phase with long-range 
   order. The line is not drawn beyond $\alpha = 0.4$ because of increasing uncertainties.}
  \label{phasediag_compact}
\end{figure}

The phase diagram for the compact $Z_4$ model with bond dissipation is shown in Fig. \ref{phasediag_compact}. It 
differs considerably from that of its noncompact counterpart, not only in the evident absence of any intermediate 
critical phase, but also in that the limit $\alpha \rightarrow 0$ is well-behaved. Here, the model is reduced to 
two uncoupled 2D Ising models, for which exact results are known and simulations are straightforward. In the 
limit of $K \rightarrow 0$ the simulations are on the other hand very difficult for the same reasons as those 
investigated by us in a similar model in Ref. \onlinecite{PhysRevB.81.104302}. Therefore, we have not strived 
to extend the phase diagram all the way down to the $\alpha$ axis in this work. Due to the qualitative difference 
in the kinetic terms for the compact and noncompact model, it is not possible to make quantitative comparison 
between the position of the phase transition line in Fig. \ref{phasediag_compact} and the two phase transition 
lines in Fig. \ref{diagramNC}.

\begin{figure}
  \centering
   \includegraphics[width=0.45\textwidth]{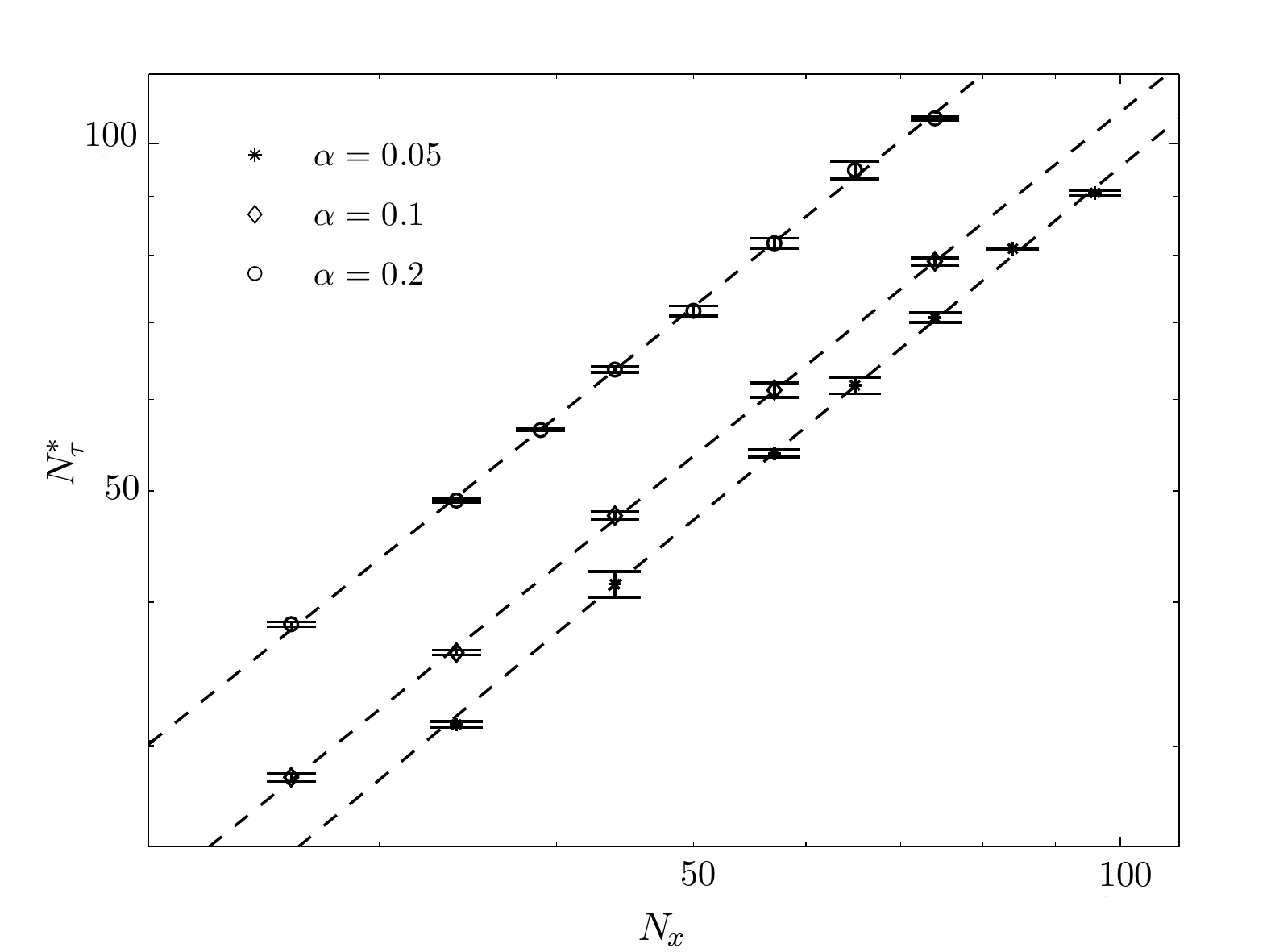}
  \caption{ Finite-size analysis of the maximum $N_\tau^\ast$ of the cumulant curves as a function of spatial system size $N_x$ used to obtain the dynamical critical exponent $z$ for the compact model. The dashed lines show the power-law fits, \cf Table \ref{tab:compact} for the results.} 
  \label{fig:Lz_vs_L_3xalpha}
\end{figure}

Turning next to the nature of the critical line in the phase diagram, we show in Fig. \ref{fig:Lz_vs_L_3xalpha} 
and Table \ref{tab:compact} results for the three 
points along the line for which we made the most effort to extract the dynamical critical exponent. These 
points are chosen so that the relative influence of the dissipation term should be qualitatively comparable 
with that for the points $(\alpha_c^{(1)},K)$ chosen for the first transition of the noncompact model. As 
for the noncompact model here and the Ising model with bond dissipation studied in Ref. 
\onlinecite{PhysRevB.81.104302}, there is no significant variation in the dynamical critical exponent from 
the expected value $z = 1$, although the tendency to greater finite-size effects for increasing $\alpha$ 
remains for both the compact and the noncompact model.

\begin{table}
\caption{Critical coupling $K_c$ and dynamical critical exponent $z$ for different values of the dissipation 
strength $\alpha$ for the compact model.}
	\begin{center}
    \begin{tabular*}{0.25\textwidth}{@{\extracolsep{\fill}}  c  c  c}
    \hline
    \hline    
    $\alpha$ & $K_c$ & $z$ \\ 
    \hline
   	0.05 	&	0.8303(4)& 1.02(2) \\
    0.1 	& 0.6753(7) & 0.99(2) \\ 
    0.2 	& 0.414(2) & 0.99(2) \\ 
    \hline
    \hline
    \end{tabular*}
    \label{tab:compact}
	\end{center}
\end{table}

\section{Discussion}
\label{Sec:discussion}

The models discussed in this paper are in some sense generalizations of the Ising spin system with bond dissipation 
discussed in Ref. \onlinecite{PhysRevB.81.104302}. For the compact $Z_4$ model the modifications come from the 
increase of the number of states from $q=2$ to $q=4$, while one for the noncompact model adds an additional
extension of the configuration space. The phase diagram for the compact model is very much like that observed for 
the dissipative Ising model,\cite{PhysRevB.81.104302} both featuring a single order-disorder phase transition line. 
The noncompact model on the other hand exhibits much richer physics in the sense that it presents, for fixed $K$ 
and $K_\tau$, two phase transitions surrounding an intermediate critical phase with power-law decaying spin 
correlations and emergent $U(1)$ symmetry. The most pressing question then pertains to the occurrence of this 
phase: Why is the discrete structure of the angle variables rendered irrelevant in a region of parameter space 
for our $Z_4$ model, when such behavior is previously known to occur only in $Z_q$ models with $q > 4$? Even 
our compactified model differs from a pure $Z_4 = Z_2 \times Z_2$ clock model, since the dissipation term couples 
the two underlying $Z_2$ models in a nontrivial way. Such models can no longer be expected \emph{a priori} to behave 
as an Ising model, and there is in principle no reason why they may not even present intermediate phases. {The 
absence of such a phase in our compact model does however indicate that we must turn to the other obvious 
difference between our model and a $Z_4$ clock model, namely that the variables in our noncompact model 
are free to drift outside the primary interval. Somehow, this added degree of freedom is enough to close the 
excitation gap.}

As observed in Fig. \ref{fig:quasi}, the underlying $Z_4$ symmetry stemming from the discreteness of the variables is 
irrelevant in the intermediate phase. Consequently, the system displays an effective continuous symmetry. Since $z=1$, 
the effective long-wavelength low-energy propagator is on a Gaussian form $1/(\omega^2+q^2)$. In addition, the system 
is effectively two dimensional due to $z=1$. In two dimensions, Gaussian fluctuations are sufficient to induce a 
critical phase given a continuous symmetry. This is analogous to the mechanism producing a critical phase in the 
classical 2D $XY$ model with an  $Z_{q>4}$ anisotropy\cite{Jose-Kadanoff-Kirkpatrick-Nelson_XY} [soft constraint 
with underlying $U(1)$ symmetry], and also for classical $Z_{q>4}$ clock models\cite{Elitzur_clock_model} (hard 
constraint). The difference in our case is that the underlying symmetry is $Z_{q=4}$.

To comment further on the origin of the critical phase, it appears that the quadratic form of the kinetic energy 
in the problem is essential for observing it. This quadratic short-range interaction term in imaginary time 
facilitates Gaussian fluctuations. Were we to use a cosine-like form of this term for noncompact variables 
(as one does for compact variables), this intermediate phase would not be found. The kinetic energy term is bounded 
from below, but not from above. Upon entering the intermediate phase from the ordered side, this term tends to suppress 
strong $\theta$ fluctuations, much more so than a kinetic term which is bounded from below {\it and} above, such 
as a cosine-like term. Only at even lower values of the dissipation are the excitation energies of larger 
$\theta$ fluctuations so low that wild $\theta$ fluctuations are possible due to the boundedness of the spatial 
coupling. At this point, the system disorders completely. If the quadratic kinetic energy term is replaced by a 
cosine-like term, wild $\theta$ fluctuations are facilitated precisely at the critical point where the $Z_4$-symmetry 
becomes irrelevant, and the system disorders directly from the $Z_4$-ordered state. Hence, for a compact model 
there will only be one phase transition separating the $Z_4$-ordered state from the completely disordered phase.

We now comment on the critical scaling between space and imaginary time in the models we have studied. For the 
compact case one has the conventional case of a critical line along which the correlation length diverges as 
$\xi \sim |K - K_c|^{-\nu}$ in space and $\xi_\tau \sim |K - K_c|^{-z\nu}$ in imaginary time, with $z$ appearing 
to remain equal to unity along the line. This picture is no longer valid in the noncompact case, as $\xi$ and 
$\xi_\tau$ are formally infinite in the entire intermediate critical phase, and $z$ can not be defined from the 
anisotropy of their divergence in this region. Furthermore, supposing that the intermediate phase shares qualities 
with the corresponding phase in classical $Z_{q>4}$ models, the correlation lengths can be expected to diverge exponentially as 
this critical phase is approached from either side as for the Kosterlitz-Thouless (KT) transition, and not as a power law as for conventional 
critical points. However, as long as the correlation length \emph{does} diverge, and this divergence is 
exponential both in space and imaginary time, the dynamical critical exponent is still well defined through 
$\xi_\tau \sim \xi^z$. Therefore, our finite size analysis is valid as $\alpha \rightarrow { \alpha_c^{(1)} }^-$ 
and $\alpha \rightarrow { \alpha_c^{(2)} }^+$ irrespective of whether these points turns out to possess
KT criticality or not. At both phase transitions we have $z=1$, signaling equally 
strong divergence of correlation lengths in space and imaginary time.

To infer from simulations on finite systems that the correlation length in fact diverges exponentially is 
exceedingly difficult,\cite{Luijten-Messingfeld,Itakura_6-state,Loison_KT_Binder} and we have not attempted 
to determine the exact nature of the phase transitions, but leave this an open question. The phase transitions 
(one or both) may be in the KT universality class, or it may belong to a class of related topological phase 
transitions.\cite{Baek-Minnhagen_5-state} This identification of the exact universality class is controversial 
even for classical clock models.\cite{PhysRevLett.96.140603,Hwang_6-state,Baek-Minnhagen-Kim_6-state} 

If we generalize the noncompact action in Sec. \ref{Sec:NC} by redefining the phase space such that the variable 
can take on all real values, Eq. (\ref{SNC}) may represent the action for a one-dimensional array of 
Josephson junctions.\cite{Chakravarty_dissipative_PT, Chakravarty_JJ_array_PRB} Recent theoretical work
\cite{Tewari-Chakravarty_dissipative_Jos, Tewari-Chakravarty_dissipative_Jos_PRB} report that such systems 
may display local quantum criticality, in the sense that the spatial coupling renormalizes to zero at the 
quantum phase transition so that the behavior is essentially $(0+1)$ dimensional. This suggests that local quantum 
criticality need not be restricted to $(2+1)$D models such as the one presented in Ref. 
~\onlinecite{Aji-Varma_orbital_currents_PRL}, but that similar unconventional criticality may be found in 
$(1+1)$D as well. Although it should be remembered that our $(1+1)$D model has discrete angle variables, 
our simulations do not show any traces of local critical behavior, in the sense that the scaling of Binder cumulants do not give $z\gg 1$.

Strictly speaking, the dynamical critical exponent is not well defined inside the intermediate phase, and the 
isotropic behavior is instead maintained by the decay exponents for the power-law spin correlation functions in 
space and time being equal. Nevertheless, for finite $N_x$ one may still assume the scaling relation 
$N_\tau^\ast = aN_x^z$ and use the ordinary procedure to extract the (effective) exponent $z$ as long as 
the system is critical, which yields $z \approx 1$ in the entire intermediate phase. We may then inspect how 
the nonuniversal prefactor $a$ changes as a reflection of  the anisotropy of the interaction in time and space. 
In the noncompact model it is possible to investigate the development of $a$ at constant $K_\tau/K$ and 
varying $\alpha$ without leaving the critical region. We find that $a$ decreases for increasing $\alpha$, 
indicating that the dissipation term contributes to making the temporal dimension less ordered than the 
spatial one. This is also in contrast with a tendency toward $(0+1)$D behavior when  
increasing the dissipation strength, as suggested in the models mentioned above.

\section{Conclusions}
\label{Sec:concl}

We have performed Monte Carlo simulations on two distinct $Z_4$-symmetric dissipative lattice models. 
In one model the phase 
variables are only defined on the interval $[0,2\pi\rangle$, while the other model has no restrictions on the variables. The 
different domains of the variables have implications for the short range interaction term in imaginary time, which again 
leads to essential differences in the behavior of the two models. The compact model features only one phase transition in 
which the $Z_4$ symmetry is spontaneously broken. On the other hand, the noncompact model displays three phases, namely 
a disordered phase with exponentially decaying spin correlations, an intermediate critical phase with quasi-long-range 
order, and finally a long-range ordered phase. 

Along the phase-transition line of the compact model, we find the dynamic critical exponent $z=1$, independent of the 
dissipation strength. In the noncompact model, we find the value $z=1$ for both phase transitions and the power-law 
decay exponents for space and imaginary time are equal in the entire phase exhibiting quasi-long-range order. 

We have shown that the issue of compactness versus noncompactness of the fundamental variables of the $Z_4$ models 
have important ramifications for their long-distance, low-energy physics.
 
\acknowledgments
The authors acknowledge useful discussions with Egil V. Herland, Mats Wallin and Henrik Enoksen. A.S. 
was supported by the Norwegian Research Council under Grant No. 167498/V30  (STORFORSK). E.B.S. and 
I.B.S thank NTNU for financial support. The work was also supported through the Norwegian consortium
for high-performance computing (NOTUR).

\appendix
\section{Quantum-to-classical mapping for compact and noncompact variables}
\label{appendix}
In this appendix we will outline the quantum-to-classical mapping for a quantum rotor model and show how the kinetic term in the resulting classical 
model depends on whether the variables are interpreted as compact or noncompact. We will first reproduce the derivation as given in Refs. 
~\onlinecite{PhysRevB.49.12115, Sondhi-Girvin_RMP} for the case of compact variables, after which we will generalize and reinterpret it for the 
noncompact case. Although there is nothing novel about this derivation, the form of the kinetic term often seems to be taken for granted in the 
literature, and a correct interpretation of the classical action in the noncompact case is crucial for our results.
As a starting point we take the (dissipationless) Hamiltonian $H_0 = T + U$ for a spatially extended system of particles, each moving on a ring.
The kinetic energy of the rotors is given by 
\begin{align}
	T = -\frac{1}{2 I}\sum_x \frac{\partial^2}{\partial \theta_x^2},
	\label{eq:T}
\end{align}
where $I$ is some inertia parameter. The (periodic) potential energy is given by Josephson-like coupling of the rotors,
\begin{align}
	U=- K\sum_x \cos(\hat{\theta}_{x+1}-\hat{\theta}_x),
	\label{eq:U}
\end{align}
with $K$ being the coupling strength.
Here we have used the angle representation where we for simplicity let $\theta$ be a continuous variable, 
and $\hat{\theta}$ is the corresponding operator.
Characteristic of a rotor model is the invariance of the system upon translations of the angle $\theta \rightarrow \theta +2\pi$. 
The eigenfunctions describing the system should therefore be $2\pi$-periodic, a requirement which immediately yields discretized 
angular momenta and energy levels.   

The partition function of the rotor system may be given by 
\begin{align}
\mathcal{Z} = \mathrm{Tr}\left(\e{-\beta(T+U)}\right).
\end{align}
We let $k_B=1$ such that $\beta$ equals inverse temperature. The trace may be evaluated by introducing a 
path integral over $M$ time slices between $\tau=0$ and $\tau=\beta$, with the width of the time 
slices given by $\Delta \tau = \beta/M$. For every time step indexed by $\tau$, we insert a complete 
set of states,
\begin{align}
	\mathcal{Z}\approx \lim_{M \to \infty} \int \mathcal{D}\theta\prod_{\tau=0}^{M-1}\langle \theta(\tau+1) | \e{-\Delta\tau T}\e{-\Delta \tau U}| \theta(\tau) \rangle.
\end{align} 
Here, $| \theta(\tau) \rangle$ is an angular eigenstate of all rotors with Trotter index $\tau$.
Since $| \theta(\tau) \rangle$ is 
an eigenstate of $\hat{\theta}$ we get
\begin{align}
\e{-\Delta\tau U}| \theta(\tau) \rangle = | \theta(\tau) \rangle \e{K\cos(\theta_{x+1,\tau}-\theta_{x,\tau})}.
\end{align}
A general matrix element describing the kinetic energy is given by
\begin{align}
T_{x,\tau} = \langle \theta_x(\tau+1) | \e{-\Delta\tau T}| \theta_x(\tau) \rangle.
\end{align}
Next, for each $\tau$ we insert a complete set of eigenstates of the kinetic energy $| n_x(\tau) \rangle$.
Because $\theta$ and $n$ are conjugate variables, we have the identity $\langle n_{x}(\tau) | \theta_x(\tau) \rangle = \exp{\lbrack {-\mathrm{i}n_{x,\tau}\theta_{x,\tau} }\rbrack}$. Inserting this, we get the general form of the matrix element for the kinetic energy
\begin{align}\label{Kinetic_element} 
T_{x,\tau} = \sum_{n_{x,\tau}} &\e{\mathrm{i}n_{x,\tau}\theta_{x,\tau+1}}\e{-\mathrm{i}n_{x,\tau}\theta_{x,\tau}}
 \e{-\frac{1}{2 I}\Delta\tau  n_{x,\tau}^2}.
\end{align}
Using the Poisson summation formula, we may write the summation over integer valued angular momenta in 
Eq. \eqref{Kinetic_element} as an integral over the continuous field $\tilde{n}$ at the cost of 
introducing another summation variable $m$:
\begin{align}
\label{villain}
T_{x,\tau} &= \sum_{m =-\infty}^{\infty}\int\mathrm{d}\tilde{n} \e{\mathrm{i}\tilde{n}(\theta_{x,\tau+1} - \theta_{x,\tau}) -\frac{1}{2 I}\Delta\tau  \tilde{n}^2}\e{2\pi\mathrm{i}m\tilde{n}} \\ \nonumber
  &=  \sum_{m=-\infty}^{\infty}\mathcal{C}\e{- \frac{I}{2 \Delta \tau} (\theta_{x,\tau+1} - \theta_{x,\tau}-2\pi m)^2} \\ \nonumber
  &\approx \mathcal{C}\e{K_\tau\cos(\theta_{x,\tau+1} - \theta_{x,\tau})},
\end{align}
where $K_\tau = \frac{I}{\Delta\tau}$, and $\mathcal{C} = \sqrt{\frac{2\pi I}{\Delta\tau}}$ is a constant 
prefactor which is henceforth dropped from the expressions. The last approximation of Eq. \eqref{villain} 
is the Villain approximation of the cosine function, which is known not to alter the universality class 
of the phase transition.

Reintroducing the matrix elements to the partition function and renaming $\Delta\tau K \to K$, we get 
\begin{align}\label{Compact_Partition} 
Z = &\int\mathcal{D}\theta\e{K_\tau\sum_\tau \sum_x\cos(\theta_{x,\tau+1}-\theta_{x,\tau})} \\ \nonumber 
	&\times\e{K \sum_\tau\sum_x \cos(\theta_{x+1,\tau}-\theta_{x,\tau})},
\end{align}
\ie, an anisotropic $XY$ model in $(1+1)$ dimensions. Note however, that we were able to cast the kinetic energy matrix 
element into the form of a sequence of Gaussians because the angular momentum eigenvalues where restricted to integer 
values. This is only the case when the canonical conjugate variable $\theta$ is restricted to a $[0,2\pi\rangle$ interval. 
In other words, the partition function given in Eq. (\ref{Compact_Partition}) reflects the interpretation of Eqs. 
\eqref{eq:T} and \eqref{eq:U} in terms of rotors. 

Equations \eqref{eq:T} and \eqref{eq:U} may also describe particles moving in an extended potential, in which case 
the state of the system after a $2\pi$ translation is distinguishable from the state prior to the translation. 
Introducing dissipation to this system by coupling $\Delta\theta$ to a bosonic bath explicitly breaks the periodicity 
of the quantum Hamiltonian, and consequently the variable $\theta$ should be treated as an extended variable from 
the outset. This necessitates a modification of the above procedure as the summation over the eigenstates in 
Eq. \eqref{Kinetic_element} has to be replaced by an integral over a continuum of momentum states. Then, the 
kinetic energy matrix element instead becomes 
\begin{align}
T_{x,\tau} &= \int\mathrm{d}n_{x,\tau}\e{\mathrm{i}n_{x,\tau}(\theta_{x,\tau+1} - \theta_{x,\tau}) - \frac{1}{2I} \Delta\tau n_{x,\tau}^2} \\ \nonumber
  &= \e{-\frac{I}{2\Delta\tau}(\theta_{x,\tau+1}- \theta_{x,\tau})^2}, 
\end{align}
where a constant factor has been ignored. Inserting this expression into the kinetic part of the partition function yields
\begin{align} 
\mathcal{Z}_\tau = &\lim_{M \to \infty}  \int \mathcal{D}\theta \e{ -\frac{I}{2}\sum_{\tau=0}^{M-1} \Delta\tau \left(\frac{\theta_{x,\tau+1}-\theta_{x,\tau}}{\Delta\tau}  \right)^2 } \\ \nonumber
									 &\equiv \int \mathcal{D}\theta \e{ -\frac{I}{2}\int_0^\beta \mathrm{d}\tau\left( \frac{\partial\theta_x}{\partial\tau}\right)^2 }.
\end{align}
This continuum expression for the action is the one 
conventionally stated in the literature both for compact and noncompact variables. However, it is always implicit 
that the imaginary time dimension is discrete by construction,\cite{negele-orland} and for most numerical computations 
it has to be treated as such in any case. One then has to choose one of two alternative discretizations of the short-range 
interaction in the imaginary time direction, depending on the interpretation of the system and the compactness of the 
variables. As shown above, the cosine-like term of Eq. \eqref{term_SC} is the natural discretization for compact 
variables, whereas the quadratic term used in Eq. \eqref{term_SNC} is associated naturally to noncompact variables.

\end{document}